\documentclass{elsart}
\usepackage{graphicx}
\usepackage{amsmath,amssymb,mathrsfs}

\begin{document}

\begin{frontmatter}

\title{Entanglement dynamics of qubits in a common environment}
\author[LZU,CKU]{Jun-Hong An} and
\ead{anjhong@lzu.edu.cn}
\author[LZU,CKU,SCU]{Shun-Jin Wang} and
\author[LZU,ITP]{Hong-Gang Luo}

\address[LZU]{Department of Modern Physics, Lanzhou University, Lanzhou 730000, P. R. China}
\address[CKU]{Center for Quantum Information Science, National Cheng Kung University,
Tainan 70101, Taiwan}
\address[SCU]{Department of Physics, Sichuan University, Chengdu 610064, P. R. China}
\address[ITP]{Institute of Theoretical Physics, Chinese Academy of
Sciences, Beijing 100080, P. R. China}
\begin{abstract}
We use the quantum jump approach to study the entanglement dynamics
of a quantum register, which is composed of two or three
dipole-dipole coupled two-level atoms, interacting with a common
environment. Our investigation of entanglement dynamics reflects
that the environment has dual actions on the entanglement of the
qubits in the model. While the environment destroys the entanglement
induced by the coherent dipole-dipole interactions, it can produce
stable entanglement between the qubits prepared initially in a
separable state. The analysis shows that it is the entangled
decoherence-free states contained as components in the initial state
that contribute to the stable entanglement. Our study indicates how
the environmental noise produces the entanglement and exposes the
interplay of environmental noise and coherent interactions of qubits
on the entanglement.
\end{abstract}


\end{frontmatter}

\section{Introduction}

Entanglement is a remarkable feature of quantum mechanics and plays a
fundamental role in quantum computation and quantum information processing.
Due to its fundamental importance for quantum information processing, it has
attracted much attention in recent years \cite{Nielsen00}. The entanglement
can be produced either by direct interactions between qubits \cite{Peter},
or by indirect interactions between the qubits through a third party \cite%
{Bose}. However, both of the above processes are constrained to the
closed system where the influence of the environment is neglected.
In real situations, a quantum system can never be isolated and will
inevitably interact with its environment. A severe effect of this
unwanted interaction is decoherence which generally leads to the
degradation of quantum coherence and entanglement. Up to now,
decoherence remains to be the main obstacle to the practical
implement of quantum information processing. Recently, the
influences of the environment on the entanglement have been
investigated extensively
\cite{Kim02,Braun02,Benatti03,Yi03,An05,Benatti05,Benatti06,Lendi07}.
It was found that in certain cases the environment can act as a
third party and be constructive to the production of the
entanglement, in contrast to its destructive role in the ordinary
situations. That is to say, depending on different coupling
mechanisms between the environment and quantum system in different
models, the environment has dual nature influenced on the
entanglement between the qubits.

In the present work, we will show that this dual influences can be
present simultaneously even in one model under certain given
coupling mechanism between the environment and system. For this
purpose, we model a quantum register by two or three two-level atoms
with dipole-dipole interactions. We study its entanglement dynamics
(i.e. the time evolution of entanglement quantity
\cite{An05,Yu02,Yu03,Yu04,Dodd04}) under the influence of a common
environment. The direct interactions between the qubits have not
been considered in the previous works
\cite{Kim02,Braun02,Benatti03,Yi03,An05}. The environment used here
is treated as a reservoir of quantum harmonic oscillators with
infinite degrees of freedom \cite{Walls94}, in contrast to the
single-mode model used by Kim et al. \cite{Kim02}. By introducing a
collective mode consisting of all the qubit modes and eliminating
the enormous irrelevant degrees of freedom of the environment, a
quantum master equation of the atomic system can be obtained under
the Born-Markovian approximation. We use the quantum jump approach
\cite{Zoller87} to study the time evolution behaviors of bi- and
tri-partite entanglement defined by the concurrence
\cite{Wootters98} and the negativity \cite{Vidal02}, respectively.
Originally, the quantum jump approach \cite{Zoller87} was proposed
to study the transition dynamics of a single atom system (for a
review, see, e.g. \cite{Plenio98}) and has been used to obtain the
operator solution of the master equation \cite{Plenio98,Plenio99}.
The merit of using quantum jump approach is that it supplies a
simple way to obtain the solution of master equation in the
multipartite system with a very large Hilbert space. The dynamical
behavior of the bi- and tri-partite entanglement shows clearly that
the environment has dual nature influenced on the entanglement in
the system studied. On the one hand, it destroys entanglement
induced by the dipole-dipole interactions, which shows the
decoherence nature of the environment. On the other hand, it induces
stable entanglement in an incoherent way, which shows the third
party role of the environment. Different from the previous works
\cite{Kim02,Braun02,Benatti03,Yi03,An05}, the dual nature of the
environment is shown simultaneously in the same system in our
situation. The stable entanglement induced by the environment is
very different from the simple oscillation behavior of entanglement
with time obtained by Kim \textit{et al.} \cite{Kim02}, where the
memory effect of the single-mode environment was considered. Here we
point out that the environment in our model as a reservoir has no
memory effect under the Markovian approximation. The environment
induces entanglement in an incoherent way since with infinite
degrees of freedom it introduces an irreversibility to the dynamics
of the qubit system.

The paper is organized as follows. In Section II we introduce the model and
the quantum jump approach. In Section III we study the entanglement dynamics
of a two-qubit system. The entanglement dynamics of a three-qubit system is
discussed in Section IV. Finally, Section V is devoted to a brief summary.

\section{Model and quantum jump approach}

We consider a quantum register composed of $N$ (=2 and 3 in the following
discussions) identical two-level atoms coupled by dipole-dipole
interactions. These atoms are assumed to be located very near, so that they
feel a common environment and interact collectively with the environment.
The environment is modeled by an electromagnetic field with infinite degrees
of freedom which is assumed at vacuum state. If the collective interaction
between atoms and the common environment is very weak and the characteristic
time of correlation function of the environment is very short compared with
the time scale of system evolution, then Born and Markovian approximations
are valid. By the standard method of quantum optics \cite{Walls94}, the
master equation of the reduced density matrix of the system can be derived
\begin{equation}
\frac{d\rho \left( t\right) }{dt}=-i[H_{I},\rho \left( t\right) ]+\frac{%
\gamma }{2}(2J_{-}\rho \left( t\right) J_{+}-J_{+}J_{-}\rho \left( t\right)
-\rho \left( t\right) J_{+}J_{-}),  \label{master}
\end{equation}%
where $\gamma $ is the decay constant of the collective mode represented by
\begin{equation*}
J_{\pm }=\sum_{i=1}^{N}\sigma _{\pm }^{i}\text{, \ \ \ \ \ }%
J_{z}=\sum_{i=1}^{N}\frac{\sigma _{z}^{i}}{2},
\end{equation*}%
and
\begin{equation*}
H_{I}=\sum_{i>j}^{N}(g_{ij}\sigma _{+}^{i}\sigma _{-}^{j}+h.c.)
\end{equation*}%
describes the dipole-dipole interactions of the qubits. The first term on
the right-hand side of Eq. (\ref{master}) generates a coherent unitary
evolution of the density matrix, while the second term represents the
decoherence effect of the environment on the system and generates an
incoherent dynamics of system. It should be noted that in general case the
strengths of dipole-dipole interactions are dependent on the positions of
the atoms. But in our model, the separation of the atoms are very small
compared to the wavelengths of the most relevant field modes, so that the
fields induce a global dipole-dipole interactions(in the long wavelength
limit), i.e. $g_{i,j}=g$. Thus the interaction Hamiltonian becomes
\begin{equation*}
H_{I}=g\left( J_{+}J_{-}-J_{z}-\frac{N}{2}\right) .
\end{equation*}%
By using individual mode representation Eq. (\ref{master}) can be rewritten
as
\begin{equation}
\frac{d\rho \left( t\right) }{dt}=-i[H_{I},\rho \left( t\right) ]+\frac{%
\gamma }{2}\sum_{i,j=1}^{N}(2\sigma _{-}^{i}\rho \left( t\right) \sigma
_{+}^{j}-\sigma _{+}^{j}\sigma _{-}^{i}\rho \left( t\right) -\rho \left(
t\right) \sigma _{+}^{j}\sigma _{-}^{i}).  \label{masterin}
\end{equation}%
It should be pointed out that the terms with $i=j$ in the sum on the
right-hand side of Eq. (\ref{masterin}) denote the individual dissipation of
each atom due to the environment, while the $i\neq j$ terms describe the
couplings between the atoms indirectly induced by the common environment.

The master equation (\ref{master}) has been used for many years to study the
superradiance effect involving the interaction of collective atomic systems
with the radiation field \cite{Dicke54,Ficek02}. The collective interaction
between the atoms and the radiative field can induce two typical atomic
states with an enhanced (the superradiant state) and a reduced (the
subradiant state) spontaneous decay rate. When the atoms are confined in a
very small region, the subradiant state is completely decoupled from the
environment, and therefore can be regarded as a decoherence-free state. A
related experiment on the superradiant spontaneous emission of two trapped
ions has been reported \cite{DeVoe96}.

The common feature of the quantum master equations is the existence of the
sandwich terms, where the reduced density matrix of the system is in between
some quantum excitation and de-excitation operators. In general, the master
equation is converted into c-number equation \cite{Walls94,Ficek02}. In this
paper, we use the quantum jump approach to obtain the exact operator
solution of the master equation (\ref{master}). As mentioned above, the
quantum-jump approach was proposed to describe single experimental
realizations of simple quantum systems (i.e. two- and three-level atoms)
\cite{Zoller87} and has been extensively used in quantum optics \cite%
{Plenio98}. By using the quantum jump approach, it is straightforward to
obtain the exact operator solution of the master equation. The advantage of
using this method is that one does not need to deal with each element of the
reduced density matrix individually, the time-dependent solution of the
reduced density matrix can be obtained directly from the formal solution
under the actions of operators.

By the quantum jump approach the reduced density matrix $\rho \left(
t\right) $ can be written as a sum of the conditional density matrices $\rho
^{(n)}(t)(n=1,2,\cdots )$ as follows
\begin{equation*}
\rho \left( t\right) =\sum_{n}\rho ^{(n)}\left( t\right) =\sum_{n}Tr_{F}\{%
\mathbb{Q}_{n}\rho _{tot}\left( t\right) \},
\end{equation*}%
where $\rho _{tot}$ is the total density matrix of the system and the
quantized environment, $Tr_{F}\{\cdot \}$ is the partial trace over the
modes of the environment, and $\mathbb{Q}_{n}$ is the projection operator on
the state of the quantized environment field that contains $n$ photons. The $%
\rho ^{(n)}\left( t\right) $ denotes the reduced density matrix of the
system with $n$ photons detected in the environment. The time evolution
equation of $\rho ^{(n)}\left( t\right) $ can be derived readily,
\begin{equation}
\frac{d\rho ^{(n)}\left( t\right) }{dt}=-i(H_{eff}\rho ^{(n)}\left( t\right)
-\rho ^{(n)}\left( t\right) H_{eff}^{\dagger })+\gamma J_{-}\rho
^{(n-1)}\left( t\right) J_{+}(1-\delta _{n0}),  \label{conequ}
\end{equation}%
where $H_{eff}=H_{I}-\frac{i\gamma }{2}J_{+}J_{-}$. Eq. (3) reflects that
the time evolution of the reduced density matrix consists of smooth
evolutions which are interrupted by instantaneous quantum jumps. It is clear
to see that the hierarchy of the conditional density matrices $\rho
^{(n)}(t) $ is terminated by instantaneous quantum jumps automatically. Due
to the dissipative effect of the environment, the effective Hamiltonian $%
H_{eff}$ governed the smooth evolution is a non-Hermitian operator. It is
straightforward to write down the formal operator solution of Eq. (\ref%
{conequ})
\begin{eqnarray}
\rho ^{(0)}(t) &=&e^{-iH_{eff}t}\rho (0)e^{iH_{eff}^{\dagger }t},  \notag \\
\rho ^{(n)}(t) &=&\gamma \int_{0}^{t}dt^{\prime }e^{-iH_{eff}(t-t^{\prime
})}J_{-}\rho ^{(n-1)}(t^{\prime })J_{+}e^{iH_{eff}^{\dagger }(t-t^{\prime
})},  \label{solu}
\end{eqnarray}%
where $\rho (0)$ is the initial state. Thus, the total reduced density
matrix of the system can be obtained by the summation over all the
conditional density operators $\rho ^{(n)}\left( t\right) $. In the
following we explicitly consider the two-qubit and three-qubit systems.

\section{Entanglement dynamics in a two-qubit system}

To simplify the following calculations, we use the collective state
representation \cite{Ficek02}. The representation space is spanned by the
common eigenstates of the complete set of commuting operators $%
\{J^{2},J_{z}\}$ for two-qubit system. This collective state representation
can also be obtained by the angular momentum additive rules and the
Clebsch-Gordan coefficients from the product state representation of the
individual ones \cite{Schiff68}. The two-qubit basis denoted by $|j,m\rangle
$ includes the spin singlet and triplet
\begin{eqnarray}
&&|0,0\rangle =\frac{1}{\sqrt{2}}(|+-\rangle -|-+\rangle )  \notag \\
&&|1,1\rangle =|++\rangle ,  \notag \\
&&|1,0\rangle =\frac{1}{\sqrt{2}}(|+-\rangle +|-+\rangle ),  \notag \\
&&|1,-1\rangle =|--\rangle ,  \label{twobas}
\end{eqnarray}%
where $|\pm\rangle$ are the eigenstates of $\sigma_{z}$ with elgenvalus $\pm1$%
, respectively. The $J_{\pm }$ act on the collective basis by
\begin{equation*}
J_{\pm }|j,m\rangle =\sqrt{j(j+1)-m(m\pm 1)}|j,m\pm 1\rangle .
\end{equation*}

When a total system is composed of two subsystems described by a
two-dimensional Hilbert state, the general measure of entanglement between
these two subsystems is the entanglement of formation \cite{Bennett96}. This
quantity can be analytically calculated as a function of the concurrence $C$
\cite{Wootters98}, which can also be taken as a measure of the entanglement.
The concurrence is defined as
\begin{equation}
C=\max (0,\lambda _{1}-\lambda _{2}-\lambda _{3}-\lambda _{4}),
\label{concur}
\end{equation}
where $\lambda _{i}(i=1,\cdots ,4)$ are eigenvalues of the matrix $[\rho
^{1/2}(\sigma _{y}\otimes \sigma _{y})\rho ^{\ast }(\sigma _{y}\otimes
\sigma _{y})\rho ^{1/2}]^{1/2}$. Here $\rho ^{\ast }$ is the complex
conjugation of $\rho $. The concurrence varies from $C=0$ for a disentangled
state to $C=1$ for a maximally entangled state. We shall use the concurrence
to quantify the degree of entanglement in the two-qubit system.

\begin{figure}[tbp]
\scalebox{0.75}{\includegraphics{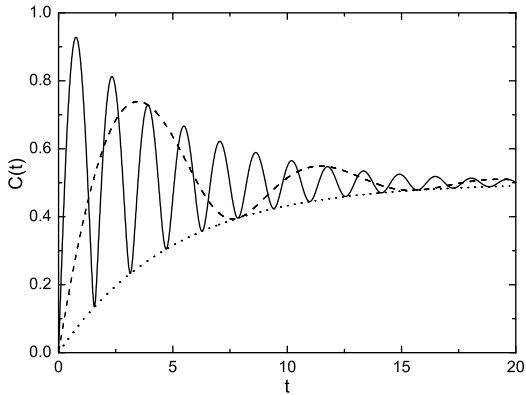}} \caption{The time
evolution of the concurrence $C(t)$. The pair of atoms is initially
prepared in the state $\protect\rho (0)=|+-\rangle \langle +-|$
for $g=1.0$ (solid line), $0.2$ (dashed line), and $0$ (dotted line) and $%
\protect\gamma =0.1$. }
\label{pureC1}
\end{figure}

In the following numerical investigations, our guide line is to explore the
following topics of entanglement dynamics:

(1) Two kinds of entanglement generation mechanisms and their
different roles played: direct coherent generation due to the atomic
dipole-dipole interaction and indirect incoherent generation due to
the common environment coupling to the qubits. From Eq.
(\ref{master}) we can see that the atomic dipole-dipole interactions
in $H_{I}$ produce a dynamical coherence which oscillates in time,
and the common environment induces an effective interaction which
produces the entanglement in steady states. Besides, the common
environment also produces a dissipative term which makes all the
eigen states of the operator of the master equation decay and
finally die out except the steady states with zero eigen values
\cite{An05}. Thus in the dissipative dynamics, the dynamical
coherence as a transit phenomenon will be destroyed finally by the
dissipation, only the environment induced coherence and entanglement
in the decoherence-free steady states survive. In the following
numerical study, the two kinds of entanglement generation show
clearly in the figures.

(2) How does the environment produce an entangled steady state from
a separable initial state? This issue is closely related to the
generation of an entanglement source and is thus very significant.

(3) How does the entanglement measure evolve and change in time from
an initial entangled state under the influence of the common
environment? This issue is related to the decoherence of a
multi-qubit system and is thus practically helpful in the
understanding of the decoherence dynamics.

(4) As the decoherence-free subspace has more than one dimensions, the
steady state will be a probabilistic mixture of the decoherence-free states.
How does one choose a proper initial state to get a desired steady state?

In order to show explicitly the entanglement dynamics of the two-qubit
system, we take the initial state $\rho (0)=|+-\rangle \langle +-|$ as our
first example. Expanding this state in the collective basis of Eqs. (\ref%
{twobas}) and substituting it into Eqs. (\ref{solu}), we can obtain
analytically the solution of the master equation (\ref{master}) after some
calculations
\begin{eqnarray}
\rho (t) &=&\frac{e^{-2\gamma t}}{2}|1,0\rangle \langle 1,0|+\frac{%
1-e^{-2\gamma t}}{2}|1,-1\rangle \langle 1,-1|  \notag \\
&&+\frac{1}{2}|0,0\rangle \langle 0,0|+\frac{e^{-i2gt-\gamma t}}{2}%
|1,0\rangle \langle 0,0|+\frac{e^{i2gt-\gamma t}}{2}|0,0\rangle \langle 1,0|.
\label{solut}
\end{eqnarray}%
The time evolution behavior of the entanglement measure, i.e. the
concurrence, can be determined by
\begin{equation}
C(t)=\frac{1}{2}\sqrt{(e^{-2\gamma t}-1)^{2}+4e^{-2\gamma t}(\sin 2gt)^{2}}.
\label{tconcu}
\end{equation}

From Eqs. (\ref{solut}) and (\ref{tconcu}) we can see that the steady
solution of the atomic system under the long time limit is
\begin{equation*}
\rho _{s}=\frac{1}{2}|1,-1\rangle \langle 1,-1|+\frac{1}{2}|0,0\rangle
\langle 0,0|,
\end{equation*}
which is an equal probabilistic mixture of the extremal states $|j,-j\rangle
$ and an entangled mixed state with $C_{s}=0.5$. The extremal states $%
|j,-j\rangle $ is decoherence-free states, as shown in Ref.
\cite{Duan98}. One of the decoherence-free states, i.e. $|0,0\rangle
$ is the so-called subradiant state \cite{Dicke54} and maximal
entangled pure state. This means that due to the symmetry of the
interaction between the system and the environment, the system
induces a collective entangled state which is immune to the
decoherence. This entangled decoherence-free state contributes to
the stable entanglement produced by the environment. From the
purification scheme proposed in Ref. \cite{Bennett96} many pairs of
this states can be used to distill out a maximal entangled state
$|0,0\rangle $ with probability 1/16.

\begin{figure}[tbp]
\scalebox{0.75}{\includegraphics{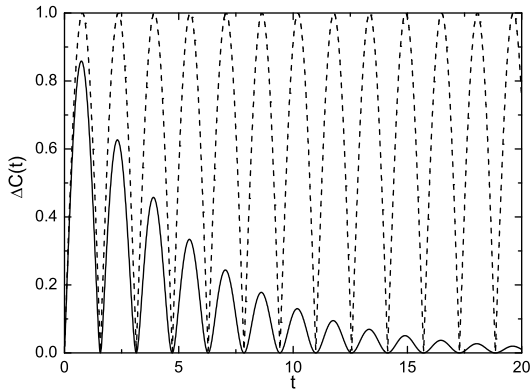}}
\caption{The time evolution of $\Delta C=C|_{g=1.0}-C|_{g=0} $ with $\protect%
\gamma =0.1$ (solid line) and $\protect\gamma =0$ (dashed line). }
\label{pureC2}
\end{figure}

The time evolution behavior of the concurrence is showed in Fig. $\ref%
{pureC1}$. We consider three cases of the dipole-dipole interaction:
absence ($g=0$, dotted line), weak ($g=0.2$, dashed line), and strong ($g=1.0$%
, solid line). In the case of $g=0$, the entanglement increases
monotonously and approaches to a stable value. Since there are no
dipole-dipole interaction between the qubits, it is quite clear that
this entanglement is induced purely by the environment. This is a
typical example that the environment plays a constructive role in
the entanglement formation between qubits. When the dipole-dipole
interaction is switched on, the entanglement induced by the
environment is not affected but an oscillating dynamical
contribution added to the total entanglement is observed. The
stronger the dipole-dipole interaction is, the faster the
entanglement oscillates, as seen easily from Eq. (\ref{tconcu}). To
see the contribution of the dipole-dipole interaction to the
entanglement, in Fig. \ref{pureC2} we plot the quantity of
$\Delta\,C = C|_{g=1.0}-C|_{g=0} $ as a function of time. It is
shown that the entanglement induced by the dipole-dipole interaction
decays rapidly as shown by solid line, which indicates that the
environment plays a destructive role to the entanglement induced by
the coherent interaction. To confirm this, we switch off the
environment, i.e., $\gamma =0$. In this situation, the entanglement
induced by the coherent interaction oscillates without dissipation,
as the dashed line shown in Fig. \ref{pureC2}. To summarize, the
environment can act as a third party to induce the entanglement
between the qubits incoherently. It also plays an usual role of
dissipation to destroy the entanglement induced coherently by the
dipole-dipole interaction. The dual nature of both construction and
destruction roles played by the environment on the entanglement
generation can be seen clearly by the above numerical analysis.

To explore the dependence of the steady state entanglement on initial
states, in the following we take the Werner states as the initial states.
The Werner states read
\begin{eqnarray}
W_{\pm } &=&(1-p)\,\frac{\mathbb{I}_{4}}{4}+p\,|{\Psi _{\pm }}\rangle
\langle {\Psi _{\pm }}|  \notag \\
&=&\frac{1+3p}{4}\,|{\Psi _{\pm }}\rangle \langle {\Psi _{\pm }}|+\frac{1-p}{%
4}(\,|{\Psi _{\mp }}\rangle \langle {\Psi _{\mp }}|+\,|{1,1}\rangle \langle
1,1|+\,|{1,-1}\rangle \langle {1,-1}|),  \label{werner}
\end{eqnarray}%
where $\mathbb{I}_{4}$ is a $4\times 4$ identity matrix, $|{\Psi _{+}}%
\rangle =|1,0\rangle $ and $|{\Psi _{-}}\rangle =|0,0\rangle $ defined by
Eq. (\ref{twobas}), and $p\in \lbrack -1/3,1]$ denotes the fidelity of $%
W_{\pm } $ to $|{\Psi _{\pm }}\rangle $. It is known that $W_{\pm }$
are entangled states for $p>1/3$ and $C_{\pm }=\frac{3p-1}{2}$. In a
similar way, one can obtain the time evolution behavior of $\rho
_{\pm }$ as
\begin{eqnarray}
&&\rho _{+}(t)=\frac{(1-p)e^{-2\gamma t}}{4}|1,1\rangle \langle 1,1|+\left\{
\frac{(1+3p)e^{-2\gamma t}}{4}+\frac{(1-p)\gamma te^{-2\gamma t}}{2}\right\}
|1,0\rangle \langle 1,0|  \notag \\
&&+\frac{1-p}{4}|0,0\rangle \langle 0,0|+\frac{(1+3p)(1-e^{-2\gamma
t})+(1-p)[2-e^{-2\gamma t}(1+2\gamma t)]}{4}|1,-1\rangle \langle 1,-1|,
\notag \\
&&\rho _{-}(t)=\frac{(1-p)e^{-2\gamma t}}{4}|1,1\rangle \langle 1,1|+\frac{%
(1-p)e^{-2\gamma t}(1+2\gamma t)}{4}|1,0\rangle \langle 1,0|  \notag \\
&&+\frac{(1-p)[3-2e^{-2\gamma t}(1+\gamma t)]}{4}|1,-1\rangle \langle 1,-1|+%
\frac{1+3p}{4}|0,0\rangle \langle 0,0|.  \label{sow}
\end{eqnarray}%
Because only the diagonal forms of collective states are involved in the
Werner states (\ref{werner}), the coherent interaction $H_{I}$ of Eq. (\ref%
{master}) has no actions on the Werner states.

\begin{figure}[tbp]
{\small \scalebox{0.6}{\includegraphics{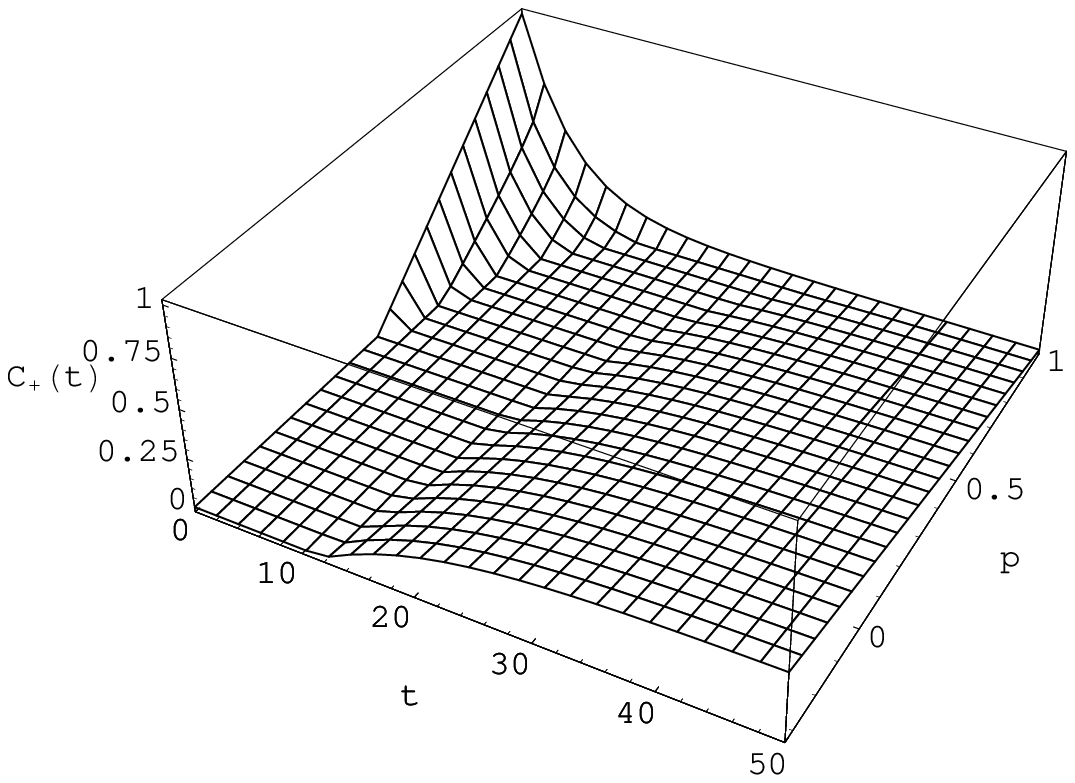}} }
\caption{The time evolution of the concurrence $C_{+}$ with different $p$.
The pair of atoms is initially prepared in the state $\protect\rho (0)=W_{+}$
and $\protect\gamma =0.1$.}
\label{mixpCP}
\end{figure}

The concurrence can be calculated in a similar way. Before discussing the
dynamical behavior of the concurrence, it is useful to consider the long
time behavior of the concurrence. With $t\rightarrow \infty $, $\rho _{\pm
}(t)$ become
\begin{eqnarray*}
&&\rho _{+s}=\frac{1-p}{4}|{\Psi _{-}}\rangle \langle {\Psi _{-}}|+\frac{3+p%
}{4}|1,-1\rangle \langle 1,-1|, \\
&&\rho _{-s}=\frac{1+3p}{4}|{\Psi _{-}}\rangle \langle {\Psi _{-}}|+\frac{%
3-3p}{4}|1,-1\rangle \langle 1,-1|,
\end{eqnarray*}%
with the steady entanglements as $C_{+s}=\frac{1-p}{4}$ and $C_{-s}=\frac{%
1+3p}{4}$, respectively. The above results show that different
initial states lead to different steady states with different
entanglements. Also from the limit case with $p=1$ one can see that
$\Psi_{+}$ is sensitive to the decoherence, while $\Psi_{-}$ is not.

\begin{figure}[tbp]
{\small \scalebox{0.6}{\includegraphics{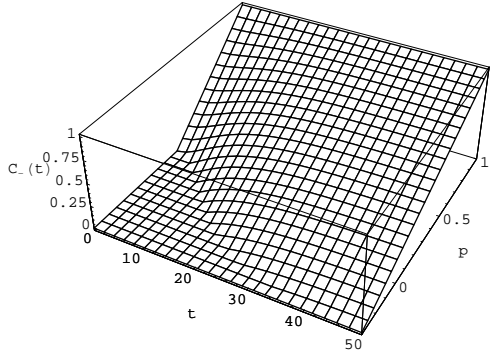}} }
\caption{The time evolution of the concurrence $C_{-}$ with different $p$.
The pair of atoms is initially prepared in the state $\protect\rho (0)=W_{-}$
and $\protect\gamma =0.1$.}
\label{mixpCM}
\end{figure}

Below we discuss entanglement dynamics in the general case. Fig. $\ref%
{mixpCP}$ and Fig. $\ref{mixpCM}$ show the concurrences of $\rho _{\pm }(t)$
as functions of $p$ and $t$. When $p<1/3$, a common characteristic of $\rho
_{\pm }(t)$ is that they have no entanglement initially. With time a stable
entanglement is formed through the environment. When $p>1/3$, the
concurrences of $\rho _{\pm }(t)$ show apparently different time evolution
behaviors. For $\rho _{+}(t)$, the entanglement contained in the initial
state $W_{+}$ is firstly destroyed by the environment due to the decoherence
effect of the environment. The entanglement experiences a \textquotedblleft
sudden death\textquotedblright \cite{Yu04}. After a finite time duration the
entanglement is reconstructed due to the constructive effect of the
environment. Physically, the initial contribution to the entanglement in $%
W_{+}$ comes from the component of $|\Psi _{+}\rangle $. With time, $|\Psi
_{+}\rangle $ (i.e. $|1,0\rangle $) approaches to the product state $%
|1,-1\rangle $ and its entanglement disappears gradually. In this process,
the decoherence-free state $|\Psi _{-}\rangle $ is unchanged and the
corresponding entanglement gradually dominates the entanglement of $\rho
_{+}(t)$ and finally, it reaches the steady state value in the long time
limit, i.e., $(1-p)/4$, as mentioned above. Different from $\rho _{+}(t)$,
the initial entanglement in $\rho _{-}(t)$ does not decay. This is because
the main contribution to the initial entanglement in this case comes from
the decoherence-free state $|\Psi _{-}\rangle $. The entanglement induced by
the environment is superposed on the initial entanglement. When $p=1$, the
Werner state reduces to the pure state $|\Psi _{-}\rangle $ which is
independent of time and free from decoherence.

From the special and general analysis above, one sees that the entanglement
of the steady state induced by the environment is rooted in the seeds of the
entangled decoherence-free states which as components have already been
contained in the initial states. This property can be used in choosing a
proper initial state to obtain a desired entangled steady state.

\section{Entanglement dynamics in a three-qubit system}

The three-qubit basis can be constructed from the addition of angular
momenta $J_{1},J_{2}$ and $J_{3}$. The representation space is spanned by
the common eigenstates of the complete set of commuting operators $%
\{J_{1,2}^{2},J^{2},J_{z}\}$. The collective basis $|j_{1,2},j,m\rangle $
can be obtained by virtue of the corresponding Clebsch-Gordan coefficients
\cite{Schiff68}
\begin{eqnarray}
&&|1,\frac{3}{2},\frac{3}{2}\rangle =|+++\rangle ,  \notag \\
&&|1,\frac{3}{2},-\frac{3}{2}\rangle =|---\rangle ,  \notag \\
&&|1,\frac{3}{2},\frac{1}{2}\rangle =\frac{1}{\sqrt{3}}(|++-\rangle
+|+-+\rangle +|-++\rangle ),  \notag \\
&&|1,\frac{3}{2},\frac{-1}{2}\rangle =\frac{1}{\sqrt{3}}(|+--\rangle
+|-+-\rangle +|--+\rangle ),  \notag \\
&&|1,\frac{1}{2},\frac{1}{2}\rangle =\frac{1}{\sqrt{6}}(2|++-\rangle
-|+-+\rangle -|-++\rangle ),  \notag \\
&&|1,\frac{1}{2},\frac{-1}{2}\rangle =\frac{1}{\sqrt{6}}(|+--\rangle
+|-+-\rangle -2|--+\rangle ),  \notag \\
&&|0,\frac{1}{2},\frac{1}{2}\rangle =\frac{1}{\sqrt{2}}(|+-+\rangle
-|-++\rangle ),  \notag \\
&&|0,\frac{1}{2},\frac{-1}{2}\rangle =\frac{1}{\sqrt{2}}(|+--\rangle
-|-+-\rangle ).  \label{bas}
\end{eqnarray}%
The $J_{\pm }$ act on the collective basis as follows
\begin{equation*}
J_{\pm }|j_{1,2},j,m\rangle =\sqrt{j(j+1)-m(m\pm 1)}|j_{1,2},j,m\pm 1\rangle
,
\end{equation*}%
which are independent of the quantum numbers $j_{1,2}$.

To discuss the entanglement dynamics in the three-qubit system, we use the
negativity proposed by Vidal and Werner to quantify the degree of
entanglement \cite{Vidal02}. The idea of this measure of the entanglement
comes from the Peres-Horodecki criterion for the separability of bipartite
systems \cite{Peres96}. The negativity was originally introduced to an
arbitrary two-qubit state $\rho $ and defined as \cite{Vidal02,Miranowicz04}
\begin{equation*}
N(\rho )=-2\sum_{i}\mu _{i}^{-}
\end{equation*}%
where $\mu _{i}^{-}$ is the negative eigenvalue of the partial transpose of $%
\rho $ with respect to the $i-$th qubit, i.e. $\rho ^{T_{i}}$. Given a
bipartite state one can calculate the partial transpose $\rho ^{T_{i}}$ of
the density operator. The state is exactly separable if $\rho ^{T_{i}}$ is
again a positive operator. However, if one of the eigenvalues of $\rho
^{T_{i}}$ is negative, the state is entangled \cite{Peres96}. From this
viewpoint, the negativity is used to quantify the degree that $\rho ^{T_{i}}$
fails to be positive and to represent the strength of quantum correlation
between the two subsystems. It was proved that the relation between $C(\rho
) $ and $N(\rho )$ is $N(\rho )\geqslant \sqrt{(1-C(\rho ))^{2}+C^{2}(\rho )}%
-[1-C(\rho )]$ for any two-qubit state \cite{Miranowicz04}. The merit of
using the negativity to quantify the entanglement is that it allows us to
investigate the entanglement properties between part $i$ and the sum of
other components in the multipartite system.

\begin{figure}[tbp]
{\small \scalebox{0.75}{\includegraphics{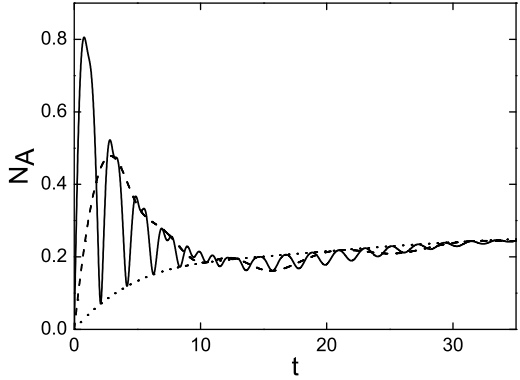}} }
\caption{The time evolution of the entanglement measured by the negativity $%
N_{A}$ when the three-atomic state is initially prepared in the state $%
\protect\rho (0)=|+-+\rangle \langle +-+|$ for $g=1.0$ (solid line), $0.2$
(dashed line), and $0$ (dotted line). The decay constant is chosen to be $%
\protect\gamma =0.1$.}
\label{NegatA}
\end{figure}

Similarly, we choose an explicit initial state, e.g., $\rho (0)=|+-+\rangle
\langle +-+|$ to discuss the entanglement dynamics in this case. Expanding
this initial state in terms of the coupled basis of Eqs. (\ref{bas})
\begin{eqnarray}
\rho \left( 0\right) &=&|\chi \rangle \langle \chi |,  \notag \\
|\chi \rangle &=&\frac{1}{\sqrt{3}}|1,\frac{3}{2},\frac{1}{2}\rangle +\sqrt{%
\frac{2}{3}}|\psi _{\frac{1}{2}}\rangle ,  \notag \\
|\psi _{\frac{1}{2}}\rangle &=&\frac{\sqrt{3}}{2}|0,\frac{1}{2},\frac{1}{2}%
\rangle -\frac{1}{2}|1,\frac{1}{2},\frac{1}{2}\rangle  \label{trein}
\end{eqnarray}%
and using the quantum jump approach, the time-dependent solution of the
master equation can be obtained analytically
\begin{eqnarray}
\rho (t) &=&\frac{e^{-4\gamma t}}{3}|1,\frac{3}{2},\frac{1}{2}\rangle
\langle 1,\frac{3}{2},\frac{1}{2}|+\frac{e^{-\gamma t}}{6}|1,\frac{1}{2},%
\frac{1}{2}\rangle \langle 1,\frac{1}{2},\frac{1}{2}|+\frac{e^{-\gamma t}}{2}%
|0,\frac{1}{2},\frac{1}{2}\rangle \langle 0,\frac{1}{2},\frac{1}{2}|  \notag
\\
&&-\frac{e^{-\frac{5\gamma t}{2}}}{\sqrt{18}}[e^{-i3gt}|1,\frac{3}{2},\frac{1%
}{2}\rangle \langle 1,\frac{1}{2},\frac{1}{2}|+e^{i3gt}|1,\frac{1}{2},\frac{1%
}{2}\rangle \langle 1,\frac{3}{2},\frac{1}{2}|]  \notag \\
&&+\frac{e^{-\frac{5\gamma t}{2}}}{\sqrt{6}}[e^{-i3gt}|1,\frac{3}{2},\frac{1%
}{2}\rangle \langle 0,\frac{1}{2},\frac{1}{2}|+e^{i3gt}|0,\frac{1}{2},\frac{1%
}{2}\rangle \langle 1,\frac{3}{2},\frac{1}{2}|]  \notag \\
&&-\frac{e^{-\gamma t}}{\sqrt{12}}[|1,\frac{1}{2},\frac{1}{2}\rangle \langle
0,\frac{1}{2},\frac{1}{2}|+|0,\frac{1}{2},\frac{1}{2}\rangle \langle 1,\frac{%
1}{2},\frac{1}{2}|]+\frac{1-e^{-\gamma t}}{2}|0,\frac{1}{2},\frac{-1}{2}%
\rangle \langle 0,\frac{1}{2},\frac{-1}{2}|  \notag \\
&&+\frac{4(e^{\gamma t}-1)e^{-4\gamma t}}{3}|1,\frac{3}{2},\frac{-1}{2}%
\rangle \langle 1,\frac{3}{2},\frac{-1}{2}|+\frac{1-e^{-\gamma t}}{6}|1,%
\frac{1}{2},\frac{-1}{2}\rangle \langle 1,\frac{1}{2},\frac{-1}{2}|  \notag
\\
&&-\frac{2e^{-\frac{5\gamma t}{2}}(e^{\gamma t}-1)}{\sqrt{18}}[e^{-i3gt}|1,%
\frac{3}{2},\frac{-1}{2}\rangle \langle 1,\frac{1}{2},\frac{-1}{2}%
|+e^{i3gt}|1,\frac{1}{2},\frac{-1}{2}\rangle \langle 1,\frac{3}{2},\frac{-1}{%
2}|]  \notag \\
&&+\frac{2e^{-\frac{5\gamma t}{2}}(e^{\gamma t}-1)}{\sqrt{6}}[e^{-i3gt}|1,%
\frac{3}{2},\frac{-1}{2}\rangle \langle 0,\frac{1}{2},\frac{-1}{2}%
|+e^{i3gt}|0,\frac{1}{2},\frac{-1}{2}\rangle \langle 1,\frac{3}{2},\frac{-1}{%
2}|]  \notag \\
&&-\frac{1-e^{-\gamma t}}{\sqrt{12}}[|1,\frac{1}{2},\frac{-1}{2}\rangle
\langle 0,\frac{1}{2},\frac{-1}{2}|+|0,\frac{1}{2},\frac{-1}{2}\rangle
\langle 1,\frac{1}{2},\frac{-1}{2}|]  \notag \\
&&+\frac{1+3e^{-4\gamma t}-4e^{-3\gamma t}}{3} |1,\frac{3}{2},\frac{-3}{2}%
\rangle \langle 1,\frac{3}{2},\frac{-3}{2}|.  \label{threeSo}
\end{eqnarray}

\begin{figure}[tbp]
{\small \scalebox{0.75}{\includegraphics{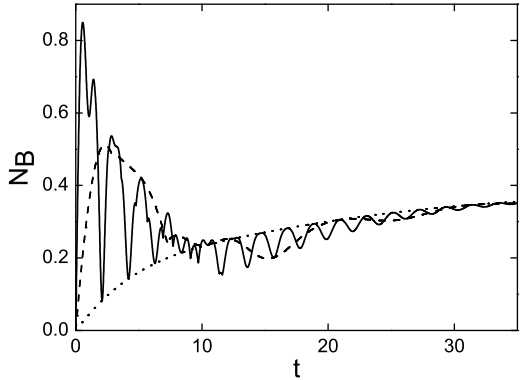}} }
\caption{The time evolution of the entanglement measured by the negativity $%
N_{B}$ when the three-atomic state is initially prepared in the state $%
\protect\rho (0)=|+-+\rangle \langle +-+|$ for $g=1.0$ (solid line), $0.2$
(dashed line), and $0$ (dotted line). The decay constant is chosen to be $%
\protect\gamma =0.1$.}
\label{NegatB}
\end{figure}

In Fig. $\ref{NegatA}$ and Fig. $\ref{NegatB}$ we show the time
evolution behavior of the negativity $N_{A}$ and $N_{B}$
corresponding to the partial transpose with respect to the first
atom $A$ and second atom $B$, respectively. In the present case,
$N_{A}=N_{C},$ where $C$ denotes the third atom. One can see that
the system shows a similar property of entanglement production as
the two-qubit system. In absence of the direct interactions, the
entanglement induced by the environment increases monotonously, then
approaches to a stable value at a large time scale for both $N_{A}$
and $N_{B}$. After the interactions are switched on, the oscillation
occurs. Moreover, during the evolution of time, this
oscillation is suppressed gradually by the environment. From Eq. ($\ref%
{threeSo}$) it is readily seen that the time-dependent solution
asymptotically tends to the steady state
\begin{eqnarray}
&&\rho _{s}=\frac{1}{3}|1,\frac{3}{2},\frac{-3}{2}\rangle \langle 1,\frac{3}{%
2},\frac{-3}{2}|+\frac{2}{3}|\psi _{\frac{-1}{2}}\rangle \langle \psi _{%
\frac{-1}{2}}|  \notag \\
&&|\psi _{\frac{-1}{2}}\rangle =\frac{\sqrt{3}}{2}|0,\frac{1}{2},\frac{-1}{2}%
\rangle -\frac{1}{2}|1,\frac{1}{2},\frac{-1}{2}\rangle ,  \label{steadt}
\end{eqnarray}%
which is a probabilistic mixture of decoherence-free state $|1,\frac{3}{2},%
\frac{-3}{2}\rangle $ and $|\psi _{\frac{-1}{2}}\rangle $ \cite{Duan98}. In
the present three-qubit case, there are two entangled decoherence-free
states which contribute to the entanglement of the steady state, i.e. $|0,%
\frac{1}{2},\frac{-1}{2}\rangle $ and $|1,\frac{1}{2},\frac{-1}{2}\rangle $.
It is interesting to note that $\rho _{s}$ is not a simple mixture of three
decoherence-free states. It still involves quantum coherence between $|0,%
\frac{1}{2},\frac{-1}{2}\rangle $ and $|1,\frac{1}{2},\frac{-1}{2}\rangle $.
By comparison of the initial state Eqs. (\ref{trein}) with the steady state
Eq. (\ref{steadt}), we see that although the jump has occurred, i.e. from $%
|\psi _{\frac{1}{2}}\rangle $ jumps to $|\psi _{\frac{-1}{2}}\rangle $ by
the action of $J_{-}$, the superposition coefficients between $|0,\frac{1}{2}%
,j_{z}\rangle $ and $|1,\frac{1}{2},j_{z}\rangle $ are not changed. This is
because that the subspaces of $|0,\frac{1}{2},j_{z}\rangle $ and $|1,\frac{1%
}{2},j_{z}\rangle $ are degenerate under the actions of both the coherent
interaction $H_{I}$ and the dissipative operators $J_{\pm }$. Anyway, the
steady state $\rho _{s}$ is still an entangled state due to the entangled
decoherence-free states contained in $|\psi _{\frac{-1}{2}}\rangle $, which
is a linear superposition of the two entangled decoherence-free states.

The above discussion in the three-qubit case further confirms the conclusion
that the environment has dual nature on the entanglement. Different from the
two-qubit case, the Hilbert space of the three-qubit system reduces into
three decoupled subspaces $|1,\frac{3}{2},j_{z}\rangle $, $|0,\frac{1}{2}%
,j_{z}\rangle $, and $|1,\frac{1}{2},j_{z}\rangle $, each of which has its
own decoherence-free state \cite{Duan98}. Only two of the decoherence-free
states are entangled. Under the time evolution, each component in the three
subspaces of the initial state approaches its corresponding decoherence-free
state. The necessary condition to induce stable entanglement is that the
initial state contains the components in the two subspaces with entangled
decoherence-free states, i.e. in the subspaces of $|0,\frac{1}{2}%
,j_{z}\rangle $ and $|1,\frac{1}{2},j_{z}\rangle $.

\section{Summary}

In conclusion, we have investigated a model of quantum register,
which is composed of two or three atoms and coupled to a common
environment. Using the quantum jump approach, the time-dependent
solution of the master equation and the entanglement dynamics of the
system are studied analytically and numerically. The dual nature of
the common environment on the entanglement and its dissipative
dynamical origin are explored in detail based on the eigen solutions
of the operator of master equation. On the one hand, due to its
induced dissipative term in the master equation, the environment
destroys the entanglement induced coherently by the atomic
dipole-dipole interactions. On the other hand, due to its induced
effective atomic interactions in the master equation, the
environment can incoherently induce the entanglement among qubits in
the decoherence-free space. The entanglement dynamics studied in the
present paper addresses the constructive role of the environment on
the entanglement production and suggests to make use of the above
positive role to construct an environment-assisted entanglement
production in quantum information processing.

\section*{Acknowledgement}

J.H.A. and S.J.W are grateful to the Center for Theoretical Sciences at
Cheng Kung University and Professor W.M. Zhang for their kind hospitality
during the visit. J.H.A. thanks the financial supports of The Fundamental
Research Fund for Physics and Mathematic of Lanzhou University under Grant
No Lzu05-02 and the NSC Grant No. 95-2816-M-006-001. The work is also
partially supported by the NNSF of China under Grants No 10604025, 10375039,
and 90503008.

\end{document}